\documentclass[12pt,thmsa]{article}
\usepackage{amsmath}
\usepackage{graphicx}
\usepackage{amsfonts}
\usepackage{amssymb}

\begin{document}
\noindent\textbf{LORENTZ\ INVARIANT\ RELATIVISTIC ELECTRODYNAMICS}

\noindent\textbf{IN THE CLIFFORD ALGEBRA FORMALISM.}

\noindent\textbf{THE FORMULATION WITH\ BIVECTOR\ FIELD\ }$F$\textbf{ }\bigskip

\qquad Tomislav Ivezi\'{c}

\qquad\textit{Ru\mbox
{\it{d}\hspace{-.15em}\rule[1.25ex]{.2em}{.04ex}\hspace{-.05em}}er Bo\v
{s}kovi\'{c} Institute, P.O.B. 180, 10002 Zagreb, Croatia}

\textit{\qquad ivezic@irb.hr\bigskip}

\qquad Received \bigskip\bigskip

\noindent In this paper we present the formulation of relativistic
electrodynamics (independent of the reference frame and of the chosen system
of coordinates in it) that uses the Faraday bivector field $F.$ This
formulation with $F$ field is a self-contained, complete and consistent
formulation that dispenses with either electric and magnetic fields or the
electromagnetic potentials. \emph{All physical quantities are defined without
reference frames} or, when some basis is introduced, every quantity is
represented as a coordinate-based geometric quantity comprising both
components and a basis. The new, observer independent, expressions for the
stress-energy vector $T(n)$ (1-vector)$,$ the energy density $U$ (scalar), the
Poynting vector $S$ and the momentum density $g$ (1-vectors), the angular
momentum density $M$ (bivector) and the Lorentz force $K$ (1-vector) are
directly derived from the field equations with $F$. The local conservation
laws are also directly derived from the field equations. \bigskip

\noindent\emph{Henceforth space by itself, and time by itself, are doomed }

\noindent\emph{to fade away into mere shadows and only a kind of union of }

\noindent\emph{the two will preserve an independent reality. H. Minkowski}\bigskip

\bigskip\bigskip

\noindent\textbf{1. INTRODUCTION}\bigskip

\noindent In the usual Clifford algebra treatments of the relativistic
electrodynamics, e.g., with multivectors $\left[  1-3\right]  $ (for a more
mathematical treatment of the Clifford algebra see also [4]), the field
equations expressed in terms of the Faraday bivector field $F$ are written as
a single equation using the $F$ field and the gradient operator $\partial$
(1-vector). In order to get the more familiar form the bivector field $F$ is
expressed (in [1,2]) in terms of the sum of a relative vector $\mathbf{E}_{H}$
(corresponds to the three-dimensional (3D) electric field vector $\mathbf{E}$)
and a relative bivector $\gamma_{5}\mathbf{B}_{H}$ ($\mathbf{B}_{H}$
corresponds to the 3D magnetic field vector $\mathbf{B}$, and $\gamma_{5}$ is
the (grade-4) pseudoscalar for the standard basis $\left\{  \gamma_{\mu
}\right\}  $) by making a space-time split in the $\gamma_{0}$ - frame, which
depends on \emph{the observer velocity }$\gamma_{0}$; the subscript 'H' is for
- Hestenes. Both $\mathbf{E}_{H}$ and $\mathbf{B}_{H}$ are, in fact,
bivectors. Similarly in [3] $F$ is decomposed in terms of 1-vector
$\mathbf{E}_{J}$ and a bivector $\mathbf{B}_{J};$ the subscript 'J' is for -
Jancewicz. It is generally accepted in the Clifford algebra formalism (and in
the tensor formalism as well) that the usual Maxwell equations with the 3D
vectors $\mathbf{E}$ and $\mathbf{B}$ and the field equations written in terms
of $F$ (or the abstract tensor $F^{ab}$ in the tensor formalism) are
completely equivalent. Further both in the tensor formalism and in the
Clifford algebra formalism it is assumed that the components of the 3D
$\mathbf{E}$ and $\mathbf{B}$ define in a unique way the components of $F$.
This means that the 3D $\mathbf{E}$ and $\mathbf{B}$ and not the $F$ field are
considered as primary quantities for the whole electromagnetism. Then in order
to get the wave theory of electromagnetism the vector potential $A$ is
introduced and $F$ is defined in terms of $A$. In that case the $F$ field
appears as the derived quantity from the potentials. Thence in all usual
treatments of the electromagnetism, both in the tensor formalism and the
Clifford algebra formalism, the theory is presented as that the $F$ field does
not have an independent existence but is defined either by the components of
the 3D $\mathbf{E}$ and $\mathbf{B}$ or by the components of the
electromagnetic potential $A$.

In all usual Clifford algebra formulations of the classical electromagnetism,
e.g., the formulations with Clifford multivectors [1-3], the standard
transformations of the 3D $\mathbf{E}$ and $\mathbf{B}$ (first derived by
Lorentz [5] and independently by Einstein [6], and subsequently quoted in
almost every textbook and paper on relativistic electrodynamics) are
considered to be the Lorentz transformations of these vectors, see [1-3]. The
same opinion holds in the tensor formalism, see [7]. These transformations are
usually derived, [7] and [1-3], by means of the above mentioned
identification, i.e., taking that the components of $F$ and of the transformed
$F^{\prime}$ are determined in the same way by the components of the 3D
$\mathbf{E}$ and $\mathbf{B}$ and of the transformed 3D $\mathbf{E}^{\prime}$
and $\mathbf{B}^{\prime}$ respectively$\mathbf{.}$ However recently important
results are achieved in the works [8] in the tensor formalism and [9] in the
Clifford algebra formalism. Namely in these works it is exactly proved that
the standard transformations of the 3D $\mathbf{E}$ and $\mathbf{B}$ \emph{are
not} relativistically correct transformations in the 4D spacetime; they are
not the Lorentz transformations of the 3D $\mathbf{E}$ and $\mathbf{B.}$ It is
also proved that the standard identification of the components of the 3D
$\mathbf{E}$ and $\mathbf{B}$ with the components of $F$ is not
relativistically correct procedure. Thence, contrary to the general belief,
the usual Maxwell equations with the 3D $\mathbf{E}$ and $\mathbf{B}$ (i.e.,
with $\mathbf{E}_{H},$ $\mathbf{B}_{H},$ or $\mathbf{E}_{J},$ $\mathbf{B}_{J}%
$) and the observer independent field equations with the $F$ field \emph{are
not} physically equivalent.

Therefore in this paper we present the formulation of relativistic
electrodynamics with the bivector field $F,$ i.e., the formulation that deal
with well-defined, observer independent, 4D quantities and not with
ill-defined quantities, the 3D $\mathbf{E}$ and $\mathbf{B,}$ or
$\mathbf{E}_{H},$ $\mathbf{B}_{H}$ (or $\mathbf{E}_{J},$ $\mathbf{B}_{J}$).
The presented formulation with the $F$ field is a self-contained, complete and
consistent formulation that does not make use either electric and magnetic
fields or the electromagnetic potential $A$ (thus dispensing with the need for
the gauge conditions). In such formulation the $F$ field is the primary
quantity for the whole classical electromagnetism. We also give the new
expressions for \emph{the observer independent} stress-energy vector $T(n)$
(1-vector)$,$ the energy density $U$ (scalar, i.e., grade-0 multivector), the
Poynting vector $S$ (1-vector)$,$ the angular momentum density $M$ (bivector)
and the Lorentz force $K$ (1-vector). They are all directly derived from the
field equations with $F$. The local charge-current density and local
energy-momentum conservation laws are also directly derived from the field
equations with $F$ and there is no need to introduce the Lagrangian and the
Noether theorem. \bigskip\medskip

\noindent\textbf{2.
SHORT\ REVIEW\ OF\ THE\ INVARIANT\ SPECIAL\ RELATIVITY\ \bigskip}

\noindent This formulation of the electromagnetism with the $F$ field
exclusively deals with well-defined 4D quantities. Namely in such formulation
physical quantities are represented by Clifford multivectors that are defined
without reference frames (when no basis has been introduced) or equivalently
by a coordinate-based geometric quantity comprising both components and a
basis (when some basis has been introduced). Thus these quantities are
independent of the chosen inertial frame of reference and of the chosen system
of coordinates in it, i.e., they are observer independent quantities. The
special relativity that exclusively uses quantities defined without reference
frames or, equivalently, the coordinate-based geometric quantities, can be
called the invariant special relativity. The reason for this name is that upon
the passive Lorentz transformations any coordinate-based geometric quantity
remains unchanged. The invariance of some 4D coordinate-based geometric
quantity upon the passive Lorentz transformations reflects the fact that such
mathematical, invariant, geometric 4D quantity represents \emph{the same
physical object} for relatively moving observers. \emph{It is taken in the
invariant special relativity that such 4D geometric quantities are
well-defined not only mathematically but also experimentally, as measurable
quantities with real physical meaning. Thus they do have an independent
physical reality. }The invariant special relativity is discussed in [10-12] in
the tensor formalism and in [13] in the Clifford algebra formalism. It is
explicitly shown in [12] that the true agreement with experiments that test
special relativity exists when the theory deals with such well-defined 4D
quantities, i.e., the quantities that are invariant upon the passive Lorentz
transformations. \emph{The principle of relativity is automatically included
in such formulation with observer independent quantities, i.e., in the
invariant special relativity, whereas in the traditional formulation of
special relativity this principle acts as the postulate established outside
the mathematical formulation of the theory. }The $F$ formulation of the
electromagnetism that is presented here is a part of the invariant special
relativity that is developed in [10-13]. In [13] we have also presented the
Clifford algebra formulations of relativistic electrodynamics with 1-vectors
$E$ and $B,$ with the real multivector $\Psi=E-e_{5}cB$ and with the complex
1-vector $\Psi=E-icB$ ($i$ is the unit imaginary). These formulations are
completely equivalent to the formulation with the $F$ field, but every of them
is an independent, consistent and complete formulation. However it is worth
noting that in these formulations of electrodynamics in [13] with $E,$ $B,$
the real and complex $\Psi$ the expressions for the stress-energy vector
$T(v)$ and all quantities derived from $T(v)$ are written for the special case
when $v,$ the velocity of observers who measure $E$ and $B$ fields is $v=cn$,
where $n$ is the unit normal to a hypersurface through which the flow of
energy-momentum ($T(n)$) is calculated. The more general case with $v\neq n$
will be reported in detail elsewhere. It is important to note that the
observer independent quantities introduced here and the field equations
written in terms of them are of the same form both in the flat and curved
spacetimes. \bigskip\medskip

\noindent\textbf{3. SHORT\ REVIEW\ OF GEOMETRIC\ ALGEBRA \bigskip}

\noindent First we provide a brief summary of geometric algebra. We write
Clifford vectors in lower case ($a$) and general multivectors (Clifford
aggregate) in upper case ($A$). The space of multivectors is graded and
multivectors containing elements of a single grade, $r$, are termed
homogeneous and written $A_{r}.$ The geometric (Clifford) product is written
by simply juxtaposing multivectors $AB$. A basic operation on multivectors is
the degree projection $\left\langle A\right\rangle _{r}$ which selects from
the multivector $A$ its $r-$ vector part ($0=$ scalar, $1=$ vector, $2=$
bivector ....). We write the scalar (grade-$0$) part simply as $\left\langle
A\right\rangle .$ The geometric product of a grade-$r$ multivector $A_{r}$
with a grade-$s$ multivector $B_{s}$ decomposes into $A_{r}B_{s}=\left\langle
AB\right\rangle _{\ r+s}+\left\langle AB\right\rangle _{\ r+s-2}%
...+\left\langle AB\right\rangle _{\ \left|  r-s\right|  }.$ The inner and
outer (or exterior) products are the lowest-grade and the highest-grade terms
respectively of the above series $A_{r}\cdot B_{s}\equiv\left\langle
AB\right\rangle _{\ \left|  r-s\right|  },$ and $A_{r}\wedge B_{s}%
\equiv\left\langle AB\right\rangle _{\ r+s}.$ For vectors $a$ and $b$ we have
$ab=a\cdot b+a\wedge b,$ where $a\cdot b\equiv(1/2)(ab+ba),$ and $a\wedge
b\equiv(1/2)(ab-ba).$ Reversion is an invariant kind of conjugation, which is
defined by $\widetilde{AB}=\widetilde{B}\widetilde{A},$ $\widetilde{a}=a,$ for
any vector $a$, and it reverses the order of vectors in any given expression.

In the treatments, e.g., $\left[  1-3\right]  $, one usualy introduces the
standard basis. The generators of the spacetime algebra (the Clifford algebra
generated by Minkowski spacetime) are taken to be four basis vectors $\left\{
\gamma_{\mu}\right\}  ,\mu=0...3,$ satisfying $\gamma_{\mu}\cdot\gamma_{\nu
}=\eta_{\mu\nu}=diag(+---).$ This basis is a right-handed orthonormal frame of
vectors in the Minkowski spacetime $M^{4}$ with $\gamma_{0}$ in the forward
light cone. The $\gamma_{k}$ ($k=1,2,3$) are spacelike vectors. This algebra
is often called the Dirac algebra $D$ and the elements of $D$ are called
$d-$numbers. The $\gamma_{\mu}$ generate by multiplication a complete basis,
the standard basis, for spacetime algebra: $1,\gamma_{\mu},\gamma_{\mu}%
\wedge\gamma_{\nu},\gamma_{\mu}\gamma_{5,}\gamma_{5}$ ($2^{4}=16$ independent
elements). $\gamma_{5}$ is the pseudoscalar for the frame $\left\{
\gamma_{\mu}\right\}  .$

We remark that the standard basis corresponds, in fact, to the specific system
of coordinates, i.e., the Einstein system of coordinates, of the chosen
inertial frame of reference. (In the Einstein system of coordinates the
Einstein synchronization $\left[  6\right]  $ of distant clocks and Cartesian
space coordinates $x^{i}$ are used in the chosen inertial frame of reference.)
However different systems of coordinates of an inertial frame of reference are
allowed and they are all equivalent in the description of physical phenomena.
For example, in $\left[  10\right]  $ two very different, but completely
equivalent systems of coordinates, the Einstein system of coordinates and
''radio'' (''r'') system of coordinates, are exposed and exploited throughout
the paper. For more detail about the ''r'' system of coordinates see, e.g.,
$\left[  10\right]  $ and references therein.

Thence instead of the standard basis $\left\{  \gamma_{\mu}\right\}  ,$
$\mu=0...3,$ for $M^{4}$ we can use some basis $\left\{  e_{\mu}\right\}  $
(the metric tensor of $M^{4}$ is then defined as $g_{\mu\nu}=e_{\mu}\cdot
e_{\nu}$) and its dual basis $\left\{  e^{\mu}\right\}  ,$ where the set of
basis vectors $e^{\mu}$ is related to the $e_{\mu}$ by the conditions $e_{\mu
}\cdot e^{\nu}=\delta_{\mu}^{\nu}$. The pseudoscalar $e_{5}$ of a frame
$\left\{  e_{\mu}\right\}  $ is defined by $e_{5}=e_{0}\wedge e_{1}%
\wedge\wedge e_{2}\wedge e_{3}.$ Then, e.g., the position 1-vector $x$ can be
decomposed in the $S$ and $S^{\prime}$ frames and in the standard basis
$\left\{  \gamma_{\mu}\right\}  $ and some non-standard basis $\left\{
e_{\mu}\right\}  $ as $x=x^{\mu}\gamma_{\mu}=x^{\mu^{\prime}}\gamma
_{\mu^{\prime}}=....=x_{e}^{\mu^{\prime}}e_{\mu^{\prime}}.$ The primed
quantities are the Lorentz transforms of the unprimed ones. Similarly any
multivector $A$ can be written as an invariant quantity with the components
and a basis, i.e., as a coordinate-based geometric quantity. In such
interpretation the Lorentz transformations are considered as passive
transformations; both the components and the basis vectors are transformed but
the whole geometric quantity remains unchanged. Thus we see that \emph{under
the passive Lorentz transformations a well-defined quantity on the 4D
spacetime, i.e., a coordinate-based geometric quantity, is an invariant
quantity.} In the usual Clifford algebra formalism, e.g., $\left[  1-4\right]
$, the Lorentz transformations are considered as active transformations; the
components of, e.g., some 1-vector relative to a given inertial frame of
reference (with the standard basis $\left\{  \gamma_{\mu}\right\}  $) are
transformed into the components of a new 1-vector relative to the same frame
(the basis $\left\{  \gamma_{\mu}\right\}  $ is not changed). (We note that a
coordinate-free form for the Lorentz transformations is presented in [13] and
it can be used both in an active way, when there is no basis, or in a passive
way, when some basis is introduced.)

The next step in the usual treatments, e.g., $\left[  1-3\right]  $, is the
introduction of a space-time split and the relative vectors. Since the usual
spacetime algebra deals exclusively with the Einstein system of coordinates it
is possible to say that a given inertial frame of reference is completely
characterized by a single future-pointing, timelike unit vector $\gamma_{0}$
($\gamma_{0}$ is tangent to the world line of an observer at rest in the
$\gamma_{0}$-system). By singling out a particular time-like direction
$\gamma_{0}$ we can get a unique mapping of spacetime into the even subalgebra
of the spacetime algebra (the Pauli subalgebra). For each spacetime point (or
event) $x$ this mapping is specified by
\begin{equation}
x\gamma_{0}=ct+\mathbf{x,\quad} ct=x\cdot\gamma_{0},\ \mathbf{x}=x\wedge
\gamma_{0}.\label{split}%
\end{equation}
To each event $x$ the equation (\ref{split}) assigns a unique time and
position in the $\gamma_{0}$-system. The set of all position vectors
$\mathbf{x}$ is the 3-dimensional position space of the observer $\gamma_{0}$
and it is designated by $P^{3}=P^{3}(\gamma_{0})=\left\{  \mathbf{x}%
=x\wedge\gamma_{0}\right\}  .$ The elements of $P^{3}$ are all spacetime
bivectors with $\gamma_{0}$ as a common factor ($x\wedge\gamma_{0}).$ They are
called \textit{the relative vectors} (relative to $\gamma_{0})$ and they will
be designated in boldface. Then a standard basis $\left\{  \mathbf{\sigma}%
_{k};k=1,2,3\right\}  $ for $P^{3},$ which corresponds to a standard basis
$\left\{  \gamma_{\mu}\right\}  $ for $M^{4}$ is given as $\mathbf{\sigma}%
_{k}=\gamma_{k}\wedge\gamma_{0}=\gamma_{k}\gamma_{0}.$ The invariant distance
$x^{2}$ then decomposes as $x^{2}=(x\gamma_{0})(\gamma_{0}x)=(ct-\mathbf{x}%
)(ct+\mathbf{x})=c^{2}t^{2}-\mathbf{x}^{2}.$ The explicit appearance of
$\gamma_{0}$ in (\ref{split}) imply that \emph{the space-time split is
observer dependent}, i.e., it is dependent on the chosen inertial frame of
reference. It has to be noted that in the Einstein system of coordinates the
space-time split of the position 1-vector $x$ (\ref{split}) gives separately
the space and time components of $x$ with their usual meaning, i.e., as in the
prerelativistic physics, and (as shown above) in the invariant distance
$x^{2}$ the spatial and temporal parts are also separated. (In the ''r''
system of coordinates there is no space-time split and also in $x^{2}$ the
spatial and temporal parts are not separated, see $\left[  10\right]  $.) This
does not mean that the Einstein system of coordinates does have some
advantages relative to other systems of coordinates and that the quantities in
the Einstein system of coordinates are more physical than, e.g., those in the
''r'' system of coordinates.

Different systems of coordinates refer to the same inertial frame of
reference, say the $S$ frame. But if we consider the geometric quantity, the
position 1-vector $x$ in another relatively moving inertial frame of reference
$S^{\prime},$ which is characterized by $\gamma_{0}^{\prime},$ then the
space-time split in $S^{\prime}$ and in the Einstein system of coordinates is
$x\gamma_{0}^{\prime}=ct^{\prime}+\mathbf{x}^{\prime}\mathbf{,}$ and this
$x\gamma_{0}^{\prime}$ is not obtained by the Lorentz transformations (or any
other coordinate transformations) from $x\gamma_{0}.$ (The hypersurface
$t^{\prime}=const.$ is not connected in any way with the hypersurface
$t=const.$) Thus the customary Clifford algebra approaches to special
relativity start with the geometric, i.e., coordinate-free, quantities, e.g.,
$x,x^{2},$ etc.$,$ which are physically well-defined. However the use of the
space-time split introduces in the customary approaches such
coordinate-dependent quantities which are not physically well-defined since
they cannot be connected by the Lorentz transformations. The main difference
between our approach to special relativity (by the use of the Clifford
algebra) and the other Clifford algebra approaches is that in our approach, as
already said, \emph{the physical meaning is attributed, both theoretically and
experimentally, only to the geometric 4D quantities, and not to their parts.
}We consider, in the same way as H. Minkowski (the motto in this paper), that
the spatial and the temporal components (e.g., $\mathbf{x}$ and $t,$
respectively) of some geometric 4D quantity (e.g., $x$) are not physically
well-defined quantities. Only their union is physically well-defined and only
such quantity does have an independent physical reality.\bigskip\medskip

\noindent\textbf{4.\ THE\ }$F$\ \textbf{FORMULATION\ OF\ ELECTRODYNAMICS
\bigskip\medskip}

\noindent\textbf{4.1. The\ Determination of the Electromagnetic Field }$F$ \bigskip

\noindent We start the exposition of electrodynamics writing the field
equations in terms of $F$ $\left[  1-3\right]  $; an electromagnetic field is
represented by a bivector-valued function $F=F(x)$ on spacetime. The source of
the field is the electromagnetic current $j$ which is a 1-vector field. Then
using that the gradient operator $\partial$ is a 1-vector field equations can
be written as a single equation
\begin{equation}
\partial F=j/\varepsilon_{0}c,\quad\partial\cdot F+\partial\wedge
F=j/\varepsilon_{0}c. \label{MEF}%
\end{equation}
The trivector part is identically zero in the absence of magnetic charge.
Notice that in $\left[  1-3\right]  $ the field equations (\ref{MEF}) are
considered to encode all of the Maxwell equations. Thus it is assumed in all
usual approaches in the Clifford algebra formalism that the field equations
(\ref{MEF}) and the usual Maxwell equations with the 3D $\mathbf{E}$ and
$\mathbf{B}$ (i.e., with $\mathbf{E}_{H},$ $\mathbf{B}_{H}$ or $\mathbf{E}%
_{J},$ $\mathbf{B}_{J}$) are physically equivalent. However, as already said,
it is exactly proved in [8] and [9] that, contrary to the general belief, such
equivalence does not exist. Thence the field equations (\ref{MEF}) are, in
fact, a relativistically correct generalization of the usual Maxwell equations
with the 3D $\mathbf{E}$ and $\mathbf{B.}$

The field bivector $F$ yields the complete description of the electromagnetic
field and, in fact, there is no need to introduce either the field vectors or
the potentials. For the given sources the Clifford algebra formalism enables
one to find in a simple way the electromagnetic field $F.$ Namely the gradient
operator $\partial$ is invertible and (\ref{MEF}) can be solved for
\begin{equation}
F=\partial^{-1}(j/\varepsilon_{0}c), \label{inef}%
\end{equation}
see, e.g., [14] and [1] Spacetime Calculus. We briefly repeat the main points
related to (\ref{inef}) from these references. The important difference is
that for us, as proved in [8] and [9], the field equations (\ref{MEF}) are not
equivalent to the usual Maxwell equations with the 3D $\mathbf{E}$ and
$\mathbf{B.}$ $\partial^{-1}$ is an integral operator which depends on
boundary conditions on $F$ and (\ref{inef}) is an integral form of the field
equations (\ref{MEF}). If the charge-current density $j(x)$ is the sole source
of $F,$ then (\ref{inef}) provides the unique solution to the field equations
(\ref{MEF}). By using Gauss' Theorem an important formula can be found that
allows to calculate $F$ at any point $y$ inside m-dimensional manifold
$\mathcal{M}$ from its derivative $\partial F$ and its values on the boundary
$\partial\mathcal{M}$ if a Green's function $G(y,x)$ is known,
\begin{equation}
F(y)=\int_{\mathcal{M}}G(y,x)\partial F(x)\mid d^{m}x\mid-\int_{\partial
\mathcal{M}}G(y,x)n^{-1}F(x)\mid d^{m-1}x\mid, \label{DF}%
\end{equation}
$n$ is a unit normal, $n^{-1}=n$ if $n^{2}=1$ or $n^{-1}=-n$ if $n^{2}=-1,$
and $G(y,x)$ is a solution to the differential equation $\partial
_{y}G(y,x)=\delta^{m}(y-x).$ ((\ref{DF}) is the relation (4.17) in [14].) If
$\partial F=0$ the first term on the right side of (\ref{DF}) vanishes. This
general relation can be applied to different examples.

An example is the determination of the expression for the classical
Li\'{e}nard-Wiechert field that is given, e.g., in [14] and [1] Spacetime
Calculus. The usual procedure ([14] and [1]) is to utilize the general
relation (\ref{DF}), in which all quantities are defined without reference
frames (Geometric calculus), and to specify it to the Minkowski spacetime
($m=4$). Then a space-time split is introduced by the relation
\begin{equation}
ct=x\cdot n=x\cdot\gamma_{0} \label{sp2}%
\end{equation}
from (\ref{split}). ($n$ in (\ref{DF}) is taken to be $\gamma_{0}$ and
(\ref{sp2}) is the equation for a 1-parameter family of spacelike hyperplanes
$S(t)$ with normal $\gamma_{0}$; $S(t)$ is a surface of simultaneous $t$.)
Further, for simplicity, $\mathcal{M}$ is taken to be the entire region
between the hyperplanes $S_{1}=S(t_{1})$ and $S_{2}=S(t_{2}).$ We note that
the same objections hold for such procedure as those that are mentioned at the
end of Sec. 3. for the space-time split. Further, as explained in the
preceding section, such procedure with the space-time split can be made only
when the metric tensor of spacetime is taken to be the Minkowski metric
tensor, i.e., when space and time are separated. For example it would not work
for the above mentioned ''r'' system of coordinates, $\left[  10\right]  $. We
shall not discuss this derivation further but we only quote the result for the
classical Li\'{e}nard-Wiechert field. The charge-current density for a
particle with charge $q$ and world line $z=z(\tau)$ with proper time $\tau$ is
$j(x)=q\int_{-\infty}^{\infty}d\tau v\delta^{4}(x-z(\tau)),$ where
$v=v(\tau)=dz/d\tau$. Then the classical Li\'{e}nard-Wiechert retarded field
for $q$ (see, e.g., Sec. 5 in [14]) is
\begin{equation}
F(x)=(q/4\pi\varepsilon_{0})\left[  r\wedge(v/c)+(1/2c^{2})r\dot
{v}(v/c)r\right]  /(r\cdot v/c)^{3}, \label{LW}%
\end{equation}
where $r=x-z$ satisfies the light-cone condition $r^{2}=0$ and $z,$ $v,$
$\dot{v}=dv/d\tau$ are all evaluated at the intersection of the backward light
cone (with vertex at $x$) and world line of that charge $q$. It is worth
noting that from the general expression (\ref{DF}) one can derive not only the
retarded interpretation for $F$ of a charge $q$ but also the advanced
interpretation and the present-time interpretation, i.e., an instantaneous
action-at-a-distance interpretation. (This present-time interpretation will be
reported elsewhere. In the tensor formalism the expressions for $F^{ab}$ and
the 4-vectors $E^{a}$ and $B^{a}$ in the present-time interpretation for an
uniform and uniformly accelerated motion of a charge $q$ are given in
[15].)\medskip\bigskip

\noindent\textbf{4.2. The\ Lorentz Force and the Motion of a Charged Particle in}

\textbf{the Electromagnetic Field }$F$ \bigskip

\noindent In the field view of particle-to-particle interaction the
electrodynamic interaction between charges is described as two-steps process;
first fields are seen as being generated from their particle sources and then
the fields so generated are perceived as interacting with some target
particle. The description of the first step in the $F$ formulation of
electrodynamics is given by the above relations (\ref{MEF}), (\ref{inef}),
(\ref{DF}) and for a particle with charge $q$ with (\ref{LW}). The second step
requires the determination of the Lorentz force in terms of $F$ and its use in
Newton's second law. This will be undertaken below.

In the Clifford algebra formalism one can easily derive the expressions for
the stress-energy vector $T(n)$ and the Lorentz force density $K$ directly
from field equations (\ref{MEF}) and from the equation for $\widetilde{F},$
the reverse of $F,$ $\widetilde{F}\widetilde{\partial}=\widetilde
{j}/\varepsilon_{0}c$ ($\widetilde{\partial}$ differentiates to the left
instead of to the right). Indeed, using (\ref{MEF}) and from the equation for
$\widetilde{F}$ one finds
\begin{equation}
T(\partial)=(-\varepsilon_{0}/2)(F\partial F)=j\cdot F/c=-K_{j},\label{TEF}%
\end{equation}
where in $(F\partial F)$ the derivative $\partial$ operates to the left and to
the right by the chain rule. The stress-energy vector $T(n)$ $\left[
1-3\right]  $ for the electromagnetic field is then defined in the $F$
formulation as
\begin{equation}
T(n)=T(n(x),x)=-(\varepsilon_{0}/2)\left\langle FnF\right\rangle
_{1}.\label{ten}%
\end{equation}
We note that $T(n)$ is a vector-valued linear function on the tangent space at
each spacetime point $x$ describing the flow of energy-momentum through a
hypersurface with normal $n=n(x)$.

The right hand side of (\ref{TEF}) yields the expression for the Lorentz force
density $K_{j},$ $K_{j}=F\cdot j/c.$ This relation shows that the Lorentz
force density $K_{j}$ can be interpreted as the rate of energy-momentum
transfer from the source $j$ to the field $F$. The Lorentz force in the $F$
formulation for a charge $q$ is $K=(q/c)F\cdot v,$ where $v$ is the velocity
1-vector of a charge $q$ (it is defined to be the tangent to its world line).

In the approaches [1,2] the Lorentz force is discussed using the space-time
split and the corresponding decomposition of $F$ into the electric and
magnetic components. Namely the bivector field $F$ is expressed in terms of
the sum of a relative vector $\mathbf{E}_{H}$ and a relative bivector
$\gamma_{5}\mathbf{B}_{H}$ by making a space-time split in the $\gamma_{0}$ -
frame
\begin{align}
F &  =\mathbf{E}_{H}+c\gamma_{5}\mathbf{B}_{H}\mathbf{,\quad E}_{H}%
=(F\cdot\gamma_{0})\gamma_{0}=(1/2)(F-\gamma_{0}F\gamma_{0}),\nonumber\\
\gamma_{5}\mathbf{B}_{H} &  =(F\wedge\gamma_{0})\gamma_{0}=(1/2c)(F+\gamma
_{0}F\gamma_{0}),\label{FB}%
\end{align}
and similarly in [3] $F$ is expressed in the $\gamma_{0}$ - frame by
$\mathbf{E}_{J}$ and $\mathbf{B}_{J}$. However the formulation with $F$ is, as
already said, a self-contained and complete formulation of electrodynamics and
there is no need for the introduction of $\mathbf{E}_{H},$ $\mathbf{B}_{H}$ or
$\mathbf{E}_{J},$ $\mathbf{B}_{J}$ from [1-3], i.e., the usual 3D electric
$\mathbf{E}$ and magnetic $\mathbf{B}$ fields. Besides it is shown in [8,9]
that the mentioned decompositions of $F$ from [1-3] lead to the standard
transformations of the 3D $\mathbf{E}$ and $\mathbf{B}$ that are not
relativistically correct transformations. (The relativistically correct
relations that connect the formulation of electrodynamics with $F$ and the
equivalent, but independent, self-contained and complete formulation with
well-defined 4D quantities, the electric and magnetic 1-vectors $E$ and $B$,
are given in [13].) Thus in the analysis of the motion of a charged particle
under the action of the Lorentz force we utilize only those parts of the usual
approaches [1-3] that are expressed only in terms of $F$ and not those
expressed by $\mathbf{E}_{H}$, $\mathbf{B}_{H}$\textbf{ }or $\mathbf{E}_{J},$
$\mathbf{B}_{J}.$ Actually we shall only quote the main results from [1-3] for
the motion of a charged particle in a constant electromagnetic field $F$ but
without using $\mathbf{E}_{H}$, $\mathbf{B}_{H}$\textbf{ }or $\mathbf{E}_{J},$
$\mathbf{B}_{J}.$

The particle equation of motion, i.e., Newton's second law is
\begin{equation}
m\dot{v}=qF\cdot v, \label{LNF}%
\end{equation}
where $\dot{v}=dv/d\tau;$ the overdot denotes differentation with respect to
proper time $\tau$. Usually [1,2] the equation (\ref{LNF}) is not solved
directly but solving the rotor equation $\dot{R}=(q/2m)FR$ and using the
invariant canonical form for $F,$ which is $F=fe^{I\varphi}=f(\cos
\varphi+I\sin\varphi);$ this holds for $F^{2}\neq0,$ and here $\gamma_{5}$ is
denoted as $I$. Let us consider that $F$ is an uniform electromagnetic field.
Then denoting ($q/m)F=\Omega$, $\Omega_{1}=f(q/m)\cos\varphi$, $\Omega
_{2}=f(q/m)\sin\varphi$, and making an invariant decomposition of the initial
velocity $v(0)$ into a component $v_{1}$ in the $f$-plane and a component
$v_{2}$ orthogonal to the $f$-plane, $v(0)=f^{-1}(f\cdot v(0))+f^{-1}(f\wedge
v(0))=v_{1}+v_{2}$, we get
\begin{equation}
v=e^{(1/2)\Omega_{1}\tau}v_{1}+e^{(1/2)\Omega_{2}\tau}v_{2}. \label{veF}%
\end{equation}
As stated in [1], Spacetime Calculus, this is an invariant decomposition of
the motion into ''electriclike'' and ''magneticlike'' components. The particle
history is obtained integrating (\ref{veF})
\begin{equation}
x(\tau)-x(0)=(e^{(1/2)\Omega_{1}\tau}-1)\Omega_{1}^{-1}v_{1}+e^{(1/2)\Omega
_{2}\tau}\Omega_{2}^{-1}v_{2}. \label{iS}%
\end{equation}
(For more details see [1-3].) This result applies for arbitrary initial
conditions and arbitrary uniform electromagnetic field $F$. Of course the
special cases can be investigated without introducing the electric and
magnetic fields. We shall not discuss here another cases that are considered
in [1-3] (e.g., a charge in an electromagnetic plane wave), but we note that
they also can be examined exclusively in terms of $F$ without introducing the
electric and magnetic fields. \medskip\bigskip

\noindent\textbf{4.3. The} \textbf{Stress-Energy Vector} $T(n)$ \textbf{and
the Quantities Derived}

\textbf{from} $T(n)$ \textbf{\bigskip}

\noindent The most important quantity for the momentum and energy of the
electromagnetic field is the observer independent stress-energy vector $T(n)$.
It can be written in the following form
\begin{equation}
T(n)=-(\varepsilon_{0}/2)\left[  (F\cdot F)n+2(F\cdot n)\cdot F\right]  .
\label{ten1}%
\end{equation}
We present a \emph{new form} for $T(n)$ (\ref{ten1}) writing it as a sum of
$n$-parallel part ($n-\parallel$) and $n$-orthogonal part ($n-\perp$)
\begin{align}
T(n)  &  =-(\varepsilon_{0}/2)\left[  (F\cdot F)+2(F\cdot n)^{2}\right]
n+\nonumber\\
&  -\varepsilon_{0}\left[  (F\cdot n)\cdot F-(F\cdot n)^{2}n\right]  .
\label{ste}%
\end{align}
The first term in (\ref{ste}) is $n-\parallel$ part and it yields the energy
density $U.$ Namely using $T(n)$ and the fact that $n\cdot T(n)$ is positive
for any timelike vector $n$ we construct the expression for \emph{the observer
independent energy density} $U$ contained in an electromagnetic field as
$U=n\cdot T(n)=\left\langle nT(n)\right\rangle ,$ (scalar, i.e., grade-0
multivector). Thus in terms of $F$ and (\ref{ste}) $U$ becomes
\begin{equation}
U=(-\varepsilon_{0}/2)\left\langle FnFn\right\rangle =-(\varepsilon
_{0}/2)\left[  (F\cdot F)+2(F\cdot n)^{2}\right]  . \label{uen1}%
\end{equation}
The second term in (\ref{ste}) is $n-\perp$ part and it is $(1/c)S$, where $S$
is \emph{the observer independent expression for the Poynting vector
}(1-vector),
\begin{equation}
S=-\varepsilon_{0}c\left[  (F\cdot n)\cdot F-(F\cdot n)^{2}n\right]  ,
\label{po1}%
\end{equation}
and, as can be seen, $n\cdot S=0$. Thus $T(n)$ expressed by $U$ and $S$ is
\begin{equation}
T(n)=Un+(1/c)S. \label{uis}%
\end{equation}
Notice that the decompositions of $T(n)$, (\ref{ten1}), (\ref{ste}) and
(\ref{uis}), are all observer independent decompositions. Further \emph{the
observer independent momentum density} $g$ is defined as $g=(1/c^{2})S$, i.e.,
$g$ is $(1/c)$ of the $n-\perp$ part from (\ref{ste})
\begin{equation}
g=-(\varepsilon_{0}/c)\left[  (F\cdot n)\cdot F-(F\cdot n)^{2}n\right]  .
\label{ge}%
\end{equation}
From $T(n)$ (\ref{ste}) one finds also the expression for \emph{the observer
independent angular-momentum density} $M$
\begin{equation}
M=(1/c)T(n)\wedge x=(1/c)U(n\wedge x)+g\wedge x. \label{em}%
\end{equation}
It has to be emphasized once again that all these definitions are the
definitions of the quantities that are independent of the chosen reference
frame and of the chosen system of coordinates in it. As I am aware they are
not presented earlier in the literature.

All these quantities can be written in some basis $\left\{  e_{\mu}\right\}
,$ which does not need to be the standard basis, as coordinate-based geometric
quantities. The field bivector $F$ can be written as $F=(1/2)F^{\alpha\beta
}e_{\alpha}\wedge e_{\beta}$ where the basis components $F^{\alpha\beta}$ are
determined as $F^{\alpha\beta}=e^{\beta}\cdot(e^{\alpha}\cdot F)=(e^{\beta
}\wedge e^{\alpha})\cdot F$. Then the quantities entering into the expressions
for $T(n),$ $U,$ $S,$ $g$ and $M$ are $F\cdot F=-(1/2)F^{\alpha\beta}%
F_{\alpha\beta},$ $F\cdot n=F^{\alpha\beta}n_{\beta}e_{\alpha},$ $(F\cdot
n)^{2}=F^{\alpha\beta}F_{\alpha\nu}n_{\beta}n^{\nu}$ and $(F\cdot n)\cdot
F=F^{\alpha\beta}F_{\alpha\nu}n_{\beta}e^{\nu}.$ Thence $T(n)$ (\ref{ten1})
becomes
\begin{equation}
T(n)=-(\varepsilon_{0}/2)\left[  (1/2)F^{\alpha\beta}F_{\beta\alpha}n^{\rho
}e_{\rho}+2F^{\alpha\beta}F_{\alpha\rho}n^{\rho}e_{\beta}\right]  ,
\label{ten2}%
\end{equation}
the energy density $U$ (\ref{uen1}) is
\begin{equation}
U=-(\varepsilon_{0}/2)\left[  (1/2)F^{\alpha\beta}F_{\beta\alpha}%
+2F^{\alpha\beta}F_{\alpha\rho}n^{\rho}n_{\beta}\right]  , \label{un2}%
\end{equation}
and the Poynting vector $S$ (\ref{po1}) becomes
\begin{equation}
S=-\varepsilon_{0}c\left[  F^{\alpha\beta}F_{\alpha\rho}n^{\rho}e_{\beta
}-F^{\alpha\beta}F_{\alpha\rho}n^{\rho}n_{\beta}n^{\lambda}e_{\lambda}\right]
. \label{po2}%
\end{equation}
In some basis $\left\{  e_{\mu}\right\}  $ we can write the stress-energy
vectors $T^{\mu}$ as $T^{\mu}=T(e^{\mu})=(-\varepsilon_{0}/2)Fe^{\mu}F.$ The
components of the $T^{\mu}$ represent the energy-momentum tensor $T^{\mu\nu}$
in the $\left\{  e_{\mu}\right\}  $ basis $T^{\mu\nu}=T^{\mu}\cdot e^{\nu
}=(-\varepsilon_{0}/2)\left\langle Fe^{\mu}Fe^{\nu}\right\rangle $, which
reduces to familiar tensor form
\begin{equation}
T^{\mu\nu}=\varepsilon_{0}\left[  F^{\mu\alpha}g_{\alpha\beta}F^{\beta\nu
}+(1/4)F^{\alpha\beta}F_{\alpha\beta}g^{\mu\nu}\right]  . \label{EMT}%
\end{equation}

In the usual Clifford algebra aproach, e.g., $\left[  1,2\right]  $, one again
makes the space-time split and considers the energy-momentum density in the
$\gamma_{0}$-system (the standard basis $\left\{  \gamma_{\mu}\right\}  $)
$T^{0}=T(\gamma^{0})=T(\gamma_{0});$ the split $T^{0}\gamma^{0}=T^{0}%
\gamma_{0}=T^{00}+\mathbf{T}^{0},$ separates $T^{0}$ into an energy density
$T^{00}=T^{0}\cdot\gamma^{0}$ and a momentum density $\mathbf{T}^{0}%
=T^{0}\wedge\gamma^{0}.$ Then from the expression for $T^{\mu}$ and the
relations (\ref{FB}) one finds [1,2] the familiar results for the energy
density $T^{00}=(\varepsilon_{0}/2)(\mathbf{E}_{H}^{2}+c^{2}\mathbf{B}_{H}%
^{2})$ and the Poyinting vector $\mathbf{T}^{0}=\varepsilon_{0}(\mathbf{E}%
_{H}\mathbf{\times} c\mathbf{B}_{H}\mathbf{),}$ where the commutator product
$A\times B$ is defined as $A\times B\equiv(1/2)(AB-BA)$. However, as already
said, the space-time split and the usual electric and magnetic fields
$\mathbf{E}_{H}$ and $\mathbf{B}_{H}$ are not only unnecessary but, as shown
in [8,9], they are relativistically incorrect. \bigskip\medskip

\noindent\textbf{4.4. The\ Local Conservation Laws in the }$F$\textbf{-
Formulation} \textbf{\bigskip}

It is well-known that from the field equations in the $F$- formulation
(\ref{MEF}) one can derive a set of conserved currents. Thus, for example, in
the $F$- formulation one derives in the standard way that $j$ from (\ref{MEF})
is a conserved current. Simply, the vector derivative $\partial$ is applied to
the field equations (\ref{MEF}) which yields
\[
(1/\varepsilon_{0}c)\partial\cdot j=\partial\cdot(\partial\cdot F).
\]
Using the identity $\partial\cdot(\partial\cdot M(x))\equiv0$ ($M(x)$ is a
multivector field) one obtains \emph{the local charge conservation law}
\begin{equation}
\partial\cdot j=0. \label{cjo}%
\end{equation}

In a like manner we find from (\ref{TEF}) that
\begin{equation}
\partial\cdot T(n)=0\label{coti}%
\end{equation}
for the free fields. This is a \emph{local energy-momentum conservation law}.
In the derivation of (\ref{TEF}) we used the fact that $T(a)$ is symmetric,
i.e., that $a\cdot T(b)=T(a)\cdot b.$ Namely using accents the expression for
$T(\partial)$ ($T(\partial)=(-\varepsilon_{0}/2)(F\partial F),$ where
$\partial$ operates to the left and to the right by the chain rule) can be
written as $T(\partial)=\acute{T}(\acute{\partial})=(-\varepsilon
_{0}/2)(\acute{F}\acute{\partial}F+F\acute{\partial}\acute{F})=0,$ since in
the absence of sources $\partial F=\acute{F}\acute{\partial}=0$ (the accent
denotes the multivector on which the derivative acts). Then from the above
mentioned symmetry of $T$ one finds that $\acute{T}(\acute{\partial})\cdot
a=\partial\cdot T(a)=0$, $\forall\ const.\ a$, which proves the equation
(\ref{coti}).

Inserting the expression (\ref{uis}) for $T(n)$ into the local energy-momentum
conservation law (\ref{coti}) we find
\begin{equation}
(n\cdot\partial)U+(1/c)\partial\cdot S=0. \label{Poy}%
\end{equation}
The relation (\ref{Poy}) is the well-known Poynting's theorem but now
completely written in terms of the observer independent quantities. Let us
introduce the standard basis $\left\{  \gamma_{\mu}\right\}  ,$ i.e., an
inertial frame of reference with the Einstein system of coordinates, and in
the $\left\{  \gamma_{\mu}\right\}  $ basis we choose that $n=\gamma_{0}$, or
in the component form it is $n^{\mu}(1,0,0,0).$ Then the familiar form of
Poynting's theorem is recovered in such coordinate system
\begin{equation}
\partial U/\partial t+\partial_{i}S^{i}=0,\qquad i=1,2,3. \label{Poy1}%
\end{equation}
It is worthwhile to note that although $U$ (\ref{uen1}) and $S$ (\ref{ste}),
taken separately, are well-defined observer independent quantities, the
relations (\ref{uis}), (\ref{coti}) and (\ref{Poy}) reveal that only $T(n)$
(\ref{uis}), as a whole quantity, i.e., the combination of $U$ and $S,$ enters
into a fundamental physical law, the local energy-momentum conservation law
(\ref{coti}). Thence one can say that only $T(n)$ (\ref{uis}), as a whole
quantity, does have a real physical meaning, or, better to say, a physically
correct interpretation. An interesting example that emphasizes this point is
the case of an uniformly accelerated charge. In the usual (3D) approach to the
electrodynamics ($\left[  7\right]  $; Jackson, Classical
Electrodynamics\textit{,} Sec. 6.8.) the Poynting vector $S$ is interpreted as
an energy flux due to the propagation of fields. In such an interpretation it
is not clear how the fields propagate along the axis of motion since for the
field points on the axis of motion one finds that $S=0$ (there is no energy
flow) but at the same time $U\neq0$ (there is an energy density). Our approach
reveals that the important quantity is $T(n)$ and not $S$ and $U$ taken
separately. $T(n)$ is $\neq0$ everywhere on the axis of motion and the local
energy-momentum conservation law (\ref{coti}) holds everywhere.

In the same way one can derive the local angular momentum conservation law,
see $\left[  1\right]  ,$ Space-Time Calculus.\bigskip\medskip

\noindent\textbf{5. COMPARISON\ WITH\ EXPERIMENTS \bigskip}

\noindent It is shown in $\left[  12\right]  $ that the usual formulation of
special relativity (which deals with the observer dependent quantities, i.e.,
the Lorentz contraction, the dilatation of time, the use of the 3D
$\mathbf{E}$ and $\mathbf{B}$, etc.,) shows only an ''apparent'' agreement
(not the true one) with the traditional and modern experiments, e.g., the
Michelson-Morley type experiments. On the contrary it is shown in $\left[
12\right]  $ that the invariant special relativity from $\left[  10\right]  $
(given in terms of \emph{geometric quantities - abstract tensors}) is in a
\emph{complete agreement} with all considered experiments. This entails that
the same complete agreement holds also for the formulation with
\emph{geometric quantities - the Clifford multivectors}, which is presented in
this paper.

In addition we briefly discuss the Trouton-Noble experiment $\left[
16\right]  $ (see also $\left[  17\right]  $). In the experiment they looked
for the turning motion of a charged parallel plate capacitor suspended at rest
in the frame of the earth in order to measure the earth's motion through the
ether. The explanations, which are given until now (see, e.g., $\left[
18\right]  $), for the null result of the experiments $\left[  16\right]  $
($\left[  17\right]  $) are not relativistically correct, since they use
quantities that are not well-defined in 4D spacetime; e.g., the Lorentz
contraction, the transformation equations for the usual 3D vectors
$\mathbf{E}$ and $\mathbf{B}$ and for the torque as the 3D vector, the
nonelectromagnetic forces of undefined nature, etc.. In our approach the
explanation is very simple and natural; the energy density $U,$ then $g$ and
$M$ and the associated integral quantities are all invariant quantities, which
means that their values are the same in the rest frame of the capacitor and in
the moving frame. Thus if there is no torque (but now as a geometric,
invariant, 4D quantity) in the rest frame then the capacitor cannot appear to
be rotating in a uniformly moving frame.\bigskip

\noindent\textbf{6. DISCUSSION\ AND\ CONCLUSIONS\bigskip}

\noindent The usual Clifford algebra approach to the relativistic
electrodynamics deals with the space-time split and the relative vectors
$\mathbf{E}_{H},$ $\mathbf{B}_{H}$ (from [1,2]) or $\mathbf{E}_{J},$
$\mathbf{B}_{J}$ (from [3]). The investigation presented in [8,9] and [13]
reveals that such approach is not relativistically correct. The usual 3D
$\mathbf{E}$ and $\mathbf{B,}$ or the relative vectors $\mathbf{E}_{H},$
$\mathbf{B}_{H}$ [1,2], or $\mathbf{E}_{J},$ $\mathbf{B}_{J}$ [3], are not
only observer dependent quantities but, as shown in [8,9], their
transformation laws are meaningless from the special relativity viewpoint;
they have nothing to do with the Lorentz transformations of the well-defined
quantities on the 4D spacetime. Here we employ quantities that are independent
of the reference frame and of the chosen system of coordinates for that frame.
We have presented the formulation of electrodynamics by means of the field
bivector $F.$ This formulation with the $F$ field is a self-contained,
complete and consistent formulation that does not make use either electric and
magnetic fields or the electromagnetic potential $A$. It provides complete and
consistent description of electromagnetic phenomena in terms of observer
independent, thus properly defined quantities on the 4D spacetime. The
formulation with the $F$ field is not physically equivalent with the usual
Maxwell formulation with the 3D vectors $\mathbf{E}$ and $\mathbf{B,}$ since,
as shown in [8,9], the transformation laws for the 3D $\mathbf{E}$ and
$\mathbf{B}$ are not relativistically correct transformations. The observer
independent field equations with the $F$ field and \emph{the new observer
independent expressions }for the stress-energy vector $T(n),$ the energy
density $U,$ the Poynting vector $S,$ the momentum density $g,$ the
angular-momentum density $M$ and the Lorentz force $K$ are presented in this
paper. The second quantization procedure, and the whole quantum
electrodynamics, will be simply constructed using geometric, invariant,
quantities $F$, $T(n)$, $U$, $S$, $g$ and $M.$ Note that the standard
covariant approaches to quantum electrodynamics, e.g., $\left[  19\right]  $,
usually deal with the component form (in the specific, i.e., the Einstein
system of coordinates) of the electromagnetic 4-potential $A$ (thus requiring
the gauge conditions too) instead of to use the geometric quantity, the
observer independent bivector field $F.$ Furthermore the standard covariant
approaches employ the definitions of the field energy and momentum, which are
not well-defined from the special relativity viewpoint. Namely, both the field
energy and momentum are defined as integrals over the \emph{three-space}, that
is, over the hypersurface $t=const.$ But the hypersurface $t=const.$ in some
reference frame $S$ cannot become (under the Lorentz transformations) the
hypersurface $t^{\prime}=const.$ in a relatively moving reference frame
$S^{\prime}.$ This is already examined for the classical electrodynamics (the
covariant formulation in the Einstein system of coordinates) by Rohrlich
$\left[  20\right]  $ and using the component form of the electric and
magnetic 4-vectors $E^{\alpha}$ and $B^{\alpha}$ (the tensor formalism) in the
first paper in $\left[  21\right]  $. (The second paper in $\left[  21\right]
$ treats relatively moving systems, e.g., a current-carrying conductor, using
the component form of the electric and magnetic 4-vectors $E^{\alpha}$ and
$B^{\alpha}.$) In this paper the local conservation laws are directly derived
from the field equations with the $F$ field and written in an invariant way.
The observer independent integral field equations and the observer independent
global conservation laws (with the definitions of the invariant field energy
and momentum) will be treated elsewhere. Particularly it has to be emphasized
that the observer independent approach to the relativistic electrodynamics
that is presented in this paper is in a complete agreement with existing
experiments that test special relativity, which is not the case with the usual
approaches. Furthermore we note that all observer independent quantities
introduced here and the field equations written in terms of them hold in the
same form both in the flat and curved spacetimes. The formalism presented here
will be the basis for the relativistically correct (without reference frames)
formulation of quantum electrodynamics and, more generally, of the quantum
field theory. \bigskip\medskip

\noindent\textbf{REFERENCES}\bigskip

\noindent1. D. Hestenes, \textit{Space-Time Algebra }(Gordon and Breach, New
York, 1966);

\textit{Space-Time Calculus; }available at: http://modelingnts.la. asu.edu/evolution.

html; \textit{New Foundations for Classical Mechanics }(Kluwer Academic

Publishers, Dordrecht, 1999) 2nd. edn..

\noindent2. S. Gull, C. Doran, and A. Lasenby, in \textit{Clifford (Geometric)
Algebras }

\textit{with Applications to Physics, Mathematics, and Engineering,} W.E. Baylis,

Ed. (Birkhauser, Boston, 1997), Chs. 6-8.; C. Doran, and A. Lasenby,

\textit{Physical Applications of Geometric Algebra,} available at: www.mrao.cam.

ac.uk/\symbol{126}Clifford/

\noindent3. B. Jancewicz, \textit{Multivectors and Clifford Algebra in Electrodynamics}

(World Scientific, Singapore, 1989).

\noindent4. D. Hestenes and G. Sobczyk, \textit{Clifford Algebra to Geometric Calculus}

(Reidel, Dordrecht, 1984).

\noindent5. H.A. Lorentz, \textit{Proceedings of the Academy of Sciences of
Amsterdam},

6 (1904), in W. Perrett and G.B. Jeffery, in \textit{The Principle of Relativity}

(Dover, New York).

\noindent6. A. Einstein, \textit{Ann. Physik.} \textbf{17}, 891 (1905), tr. by
W. Perrett and G.B.

Jeffery, in \textit{The Principle of Relativity} (Dover, New York).

\noindent7. J.D. Jackson, \textit{Classical Electrodynamics} (Wiley, New York,
1977) 2nd

edn.; L.D. Landau and E.M. Lifshitz, \textit{The Classical Theory of Fields,}

(Pergamon, Oxford, 1979) 4th edn.; C.W. Misner, K.S.Thorne, and J.A.

Wheeler, \textit{Gravitation} (Freeman, San Francisco, 1970).

\noindent8. T. Ivezi\'{c}, \textit{hep-th}/0302188; to be published in
\textit{Found. Phys.} \textbf{33}, (issue

9) (2003).

\noindent9. T. Ivezi\'{c}, \textit{physics/0304085}.

\noindent10. T. Ivezi\'{c}, \textit{Found. Phys.} \textbf{31}, 1139 (2001).

\noindent11. T. Ivezi\'{c}, \textit{Annales de la Fondation Louis de Broglie}
\textbf{27}, 287 (2002).

\noindent12. T. Ivezi\'{c}, \textit{Found. Phys. Lett.} \textbf{15}, 27
(2002); \textit{physics}/0103026; \textit{physics}/

0101091.

\noindent13. T. Ivezi\'{c}, \textit{hep-th}/0207250; \textit{hep-ph}/0205277.

\noindent14. D. Hestenes, in \textit{Clifford Algebras and their Applications
in }

\textit{Mathematical Physics}, F. Brackx et al, Eds. (Kluwer Academic

Publishers, Dordrecht, 1993).

\noindent15. T. Ivezi\'{c} and Lj. \v{S}kovrlj, unpublished results; Lj. \v
{S}kovrlj, \textit{Thesis} (2002)

(in Croatian).

\noindent16. F.T. Trouton and H.R. Noble, \textit{Philos. Trans. R. Soc.
London Ser. A}

\textbf{202}, 165 (1903).

\noindent17. H.C. Hayden, \textit{Rev. Sci. Instrum.} \textbf{65}, 788 (1994).

\noindent18. A.K. Singal, \textit{J. Phys. A: Math. Gen.} \textbf{25} 1605
(1992); \textit{Am. J. Phys.} \textbf{61},

428 (1993); S. A. Teukolsky, \textit{Am. J. Phys.} \textbf{64}, 1104 (1996); O.D.

Jefimenko, \textit{J. Phys. A: Math. Gen.} \textbf{32,} 3755 (1999).

\noindent19. J.D. Bjorken and S.D. Drell, \textit{Relativistic Quantum Field} (McGraw-Hill,

New York, 1964); F. Mandl and G. Shaw, \textit{Quantum Field Theory} (John

Wiley \&Sons, New York, 1995); S. Weinberg, \textit{The} \textit{Quantum
Theory of}

\textit{Fields, Vol. I Foundations, }(Cambridge University Press, Cambridge,

1995).

\noindent20. F. Rohrlich, \textit{Phys. Rev. D} \textbf{25}, 3251 (1982).

\noindent21. T. Ivezi\'{c}, \textit{Found. Phys. Lett.} \textbf{12}, 105
(1999); \textit{Found. Phys. Lett}. \textbf{12},

507 (1999).
\end{document}